\documentclass[twocolumn,prl,showpacs,preprintnumbers,amsmath,amssymb]{revtex4}

\usepackage{graphicx}% Include figure files
\usepackage{dcolumn}% Align table columns on decimal point
\usepackage{bm}% bold math

\begin{document}

\title{Phonon Transport in Suspended Single Layer Graphene}

\author {Xiangfan Xu$^{1,2}$, Yu Wang$^{3}$, Kaiwen Zhang$^{1,4}$, Xiangming Zhao$^{1,4}$, Sukang Bae$^{5}$, Martin Heinrich$^{6}$,  Cong Tinh Bui$^{6}$, Rongguo Xie$^{1,4,7}$, John T. L. Thong$^{7,6}$, Byung Hee Hong$^{5,8}$, Kian Ping Loh$^{3,6}$, Baowen
Li$^{1,4,6}$, Barbaros \"Ozyilmaz$^{1,2,6}$}
\email{barbaros@nus.edu.sg}

\affiliation{$^{1}$Department of Physics, National University of
Singapore, Singapore 117542}

\affiliation{$^{2}$NanoCore, 4 Engineering Drive 3, National
University of Singapore, Singapore 117576}

\affiliation{$^{3}$Department of Chemistry, National University of
Singapore, Singapore 117543}

\affiliation{$^{4}$Centre for Computational Science and
Engineering, National University of Singapore, Singapore 117542}

\affiliation{$^{5}$SKKU Advanced Institute of Nanotechnology
(SAINT) and Center for Human Interface Nano Technology (HINT),
Sungkyunkwan University, Suwon 440-746, Korea}

\affiliation{$^{6}$NUS Graduate School for Integrative Science and
Engineering, Singapore 117456}

\affiliation{$^{7}$ Department of Electrical and Computer
Engineering, National University of Singapore, Singapore 117576}

\affiliation{$^{8}$Department of Chemistry, Sungkyunkwan
University, Suwon 440-746, Korea}

\date{\today}% It is always \today, today,

\begin{abstract}
We report the first temperature dependent phonon transport
measurements in suspended Cu-CVD single layer graphene (SLG) from
15K to 380K using microfabricated suspended devices. The thermal
conductance per unit cross section $\sigma$/A increases with
temperature and exhibits a peak near $T$ $\sim$ 280K ($\pm$10K)
due to the Umklapp process. At low temperatures ($T$$<$140K), the
temperature dependent thermal conductivity scales as
$\sim$$T$$^{1.5}$, suggesting that the main contribution to
thermal conductance arises from flexural acoustic (ZA) phonons in
suspended SLG. The $\sigma$/A reaches a high value of
1.7$\times$10$^5$ $T^{1.5}$ W/m$^2$K, which is approaching the
expected ballistic phonon thermal conductance for two-dimensional
graphene sheets. Our results not only clarify the ambiguity in the
thermal conductance, but also demonstrate the potential of Cu-CVD
graphene for heat related applications.

\end{abstract}

% thermal in graphene, phonons in graphene, graphene, Nanoscale materials and structures: fabrication and characterization

\maketitle

Heat conduction in low-dimensional systems has been extensively
studied in past decades because of its importance in understanding
the microscopic mechanism of thermal transport phenomena, and its
potential applications in manipulating and controlling heat flow
\cite{lowD}.  Although significant progress has been made for
one-dimensional systems \cite{lowD}, the study of heat conduction
in two-dimensional (2D) systems is still in its infancy due to the
lack of proper materials. The discovery of graphene has changed
this scenario \cite{graphene,neto}. It not only provides us with
an ideal platform for thermal transport studies in 2D systems, but
also opens the door for many novel heat related applications
\cite{pop}. However, difficulties in integrating graphene sheets
with device structures needed for measuring intrinsic thermal
properties have resulted in experiments which only offer partial
answers. Pioneering Raman based measurements at room temperature
have indicated exceptional high thermal conductivity in graphene
\cite{balandin1,balandin2,balandin3}, far exceeding that of
diamond \cite{diamond} and carbon nanotubes \cite{Nanotube,CNT}.
However, recent measurements have shown that these values can
easily vary by one order of magnitude from 600 W/mK to 5300 W/mK
\cite{balandin2,Geim,Ruoff}. Efforts to develop traditional
suspended microfabricated heater wires have been only successful
by compromising on the SiO$_2$ substrate, resulting in significant
phonon leakage into the substrate and strong interfacial
scattering between graphene and the underlying substrate
\cite{LSScience}. Therefore, the temperature dependent thermal
conductivity $\kappa$($T$) needs to be investigated in suspended
graphene. This will be the key in understanding the 2D nature of
phonons in graphene and is crucial in identifying which phonon
branch dominating the heat conduction.

\begin{figure}[tbp]
\includegraphics[width=7.5cm]{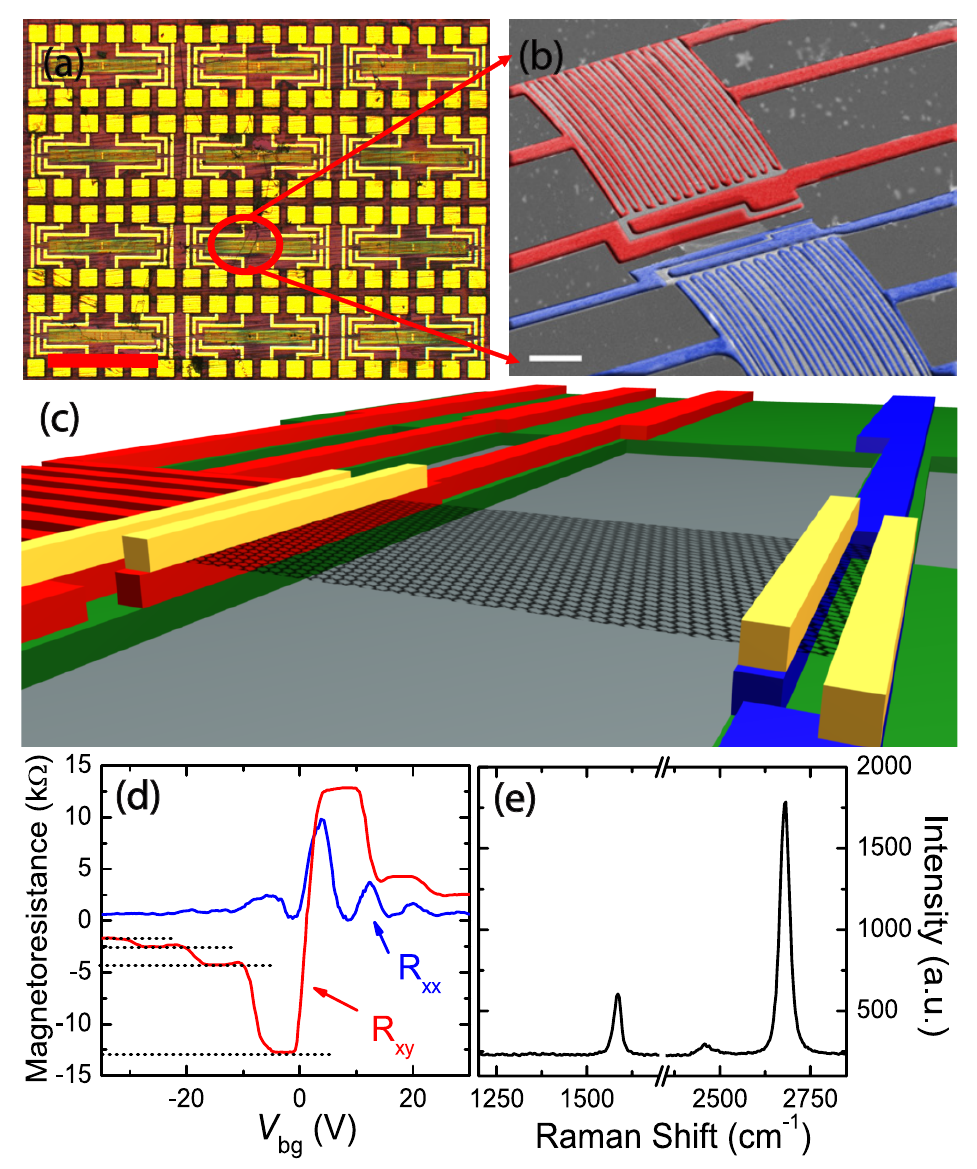}
\caption{(a)Microdevice array after CVD graphene transfer and
before suspension. Only a 3$\times$4 array section is shown, scale
bar is 1mm. (b)False-color SEM image in the center of the
suspended device. The red and blue Pt coils are the heater
(R$_{h}$) and sensor (R$_{s}$), respectively. They are thermally
connected by suspended graphene (grey rectangle in the middle).
The inner four Pt electrodes are connected to graphene, but
electrically isolated from Pt coils. The scale bar is 5$\mu$m.
(c)Schematics of a graphene sheet clamped between Cr/Au (yellow)
and Pt electrodes (blue and red) on SiN$_x$ membranes (green).
Lateral dimensions are not to scale. (d)Quantum Hall effect at
$T$=2K, $B$=9T and (e)Raman spectrum of the same batch of graphene
transferred on regular Si/SiO$_2$ wafers.}
\end{figure}

In this Letter, we present the first temperature dependent thermal
transport measurements in a suspended single layer graphene (SLG)
sheet in the temperature range from 15K to 380K. Our approach
takes advantage of the recent progress in wafer-scale graphene
growth by the Cu-based chemical vapor deposition (Cu-CVD) method
\cite{CVDruoff,CVDKorea}. The thermal conductivity follows a
$T$$^{1.5}$ power law, demonstrating that most of the heat in free
standing graphene is carried by flexural acoustic (ZA) phonons. At
room temperature, the thermal conductance per unit cross section
$\sigma$/A in Cu-CVD graphene reaches a high value of
3.8$\times$10$^8$ W/m$^2$K. Below 140K, this high value is found
to be approaching the theoretical ballistic limit in graphene
sheets \cite{ballisticMingo,ballisticNL}.

We employed typical prepatterned suspended heater wires to measure
the thermal conductance and thermal power in suspended single
layer graphene. Similar methods have been used to study thermal
transport in carbon nanotubes \cite{Nanotube,Nanotube2} and
nanowire \cite{Nanowire}. The structures were fabricated on
silicon nitride (SiN$_x$)/Si multilayer wafers and the fabrication
details have already been discussed by Shi $et$ $al$. \cite{JHT}.
The transfer of exfoliated graphene onto such fragile structures
is difficult and remains very challenging. Fortunately, the recent
progress in Cu-CVD graphene allows us to overcome many of these
challenges by the sheer number of available junctions on a single
chip. The growth of SLG by Cu-CVD method is described elsewhere in
detail \cite{CVDKorea}. The outstanding electronic properties and
sub-100 micrometers grain size \cite{grainsize} suggest that the
physical properties of Cu-CVD graphene are comparable to that of
exfoliated graphene. Graphene films were transferred onto these
prepatterned structures before suspending them (Figure 1a)
\cite{supplementary}. They were then patterned into
micrometer-size rectangular structures by standard e-beam
lithography, followed by an O$_{2}$ plasma. In a second
lithography step, 30nm Cr/Au bars were deposited on both ends of
graphene to ensure good thermal contact with the Pt electrodes
underneath graphene (Figure 1c). These Cr/Au bars also securely
clamp graphene onto the Pt electrodes during the subsequent
fabricating steps. After suspending graphene by etching in KOH,
the devices were dried using a critical point dryer to avoid
damage due to surface tension.

The scanning electron microscope image of a typical
microfabricated device is shown in Figure 1b. The suspended device
consists of two 25$\times$20 $\mu$m$^2$ SiN$_x$ membranes, each of
which is supported by six 400$\mu$m-long, 0.3$\mu$m-thick and
1.5$\mu$m-wide suspended SiN$_x$ beams. A 60nm-thick platinum
resistor coil is deposited on top of each SiN$_x$ membrane. Since
the resistance of the Pt resistors changes with temperature, they
can serve as both heater (R$_{h}$) and temperature sensor
(R$_{s}$). The suspended graphene (grey sheet in the middle of
Figure 1b) bridges the two SiN$_x$ membranes and is electrically
isolated from the Pt coils. In total, three samples from two
separate Cu-CVD growth runs have been measured. Here we discuss
representative data based on one graphene sheet, which is
patterned to a width of 3$\mu$m and a suspended length of 500nm.
Since there is no back gate for such suspended graphene samples
(etch depth$\approx$200$\mu$m), we could neither vary the carrier
density nor calculate the mobility. However, this is not a severe
limitation in our measurements due to the following reasons: a)
thermal conductance in graphene is weakly dependent on carrier
density; b) we can set the Fermi level near Dirac point by
$in$-$situ$ annealing in vacuum \cite{suspended}; c) the sample
quality can be investigated by characterizing the same batch of
graphene on SiO$_2$. The measured mobility is approximately
7700cm$^{-2}$/Vs at $T$=2K, which is comparable to that in
exfoliated graphene. The half-integer quantum Hall measurements at
$T$=2K and $B$=9T (Figure 1d) and Raman measurements (Figure 1e)
also indicate high quality of our graphene samples.

\begin{figure}[tbp]
\includegraphics[width=7.5cm]{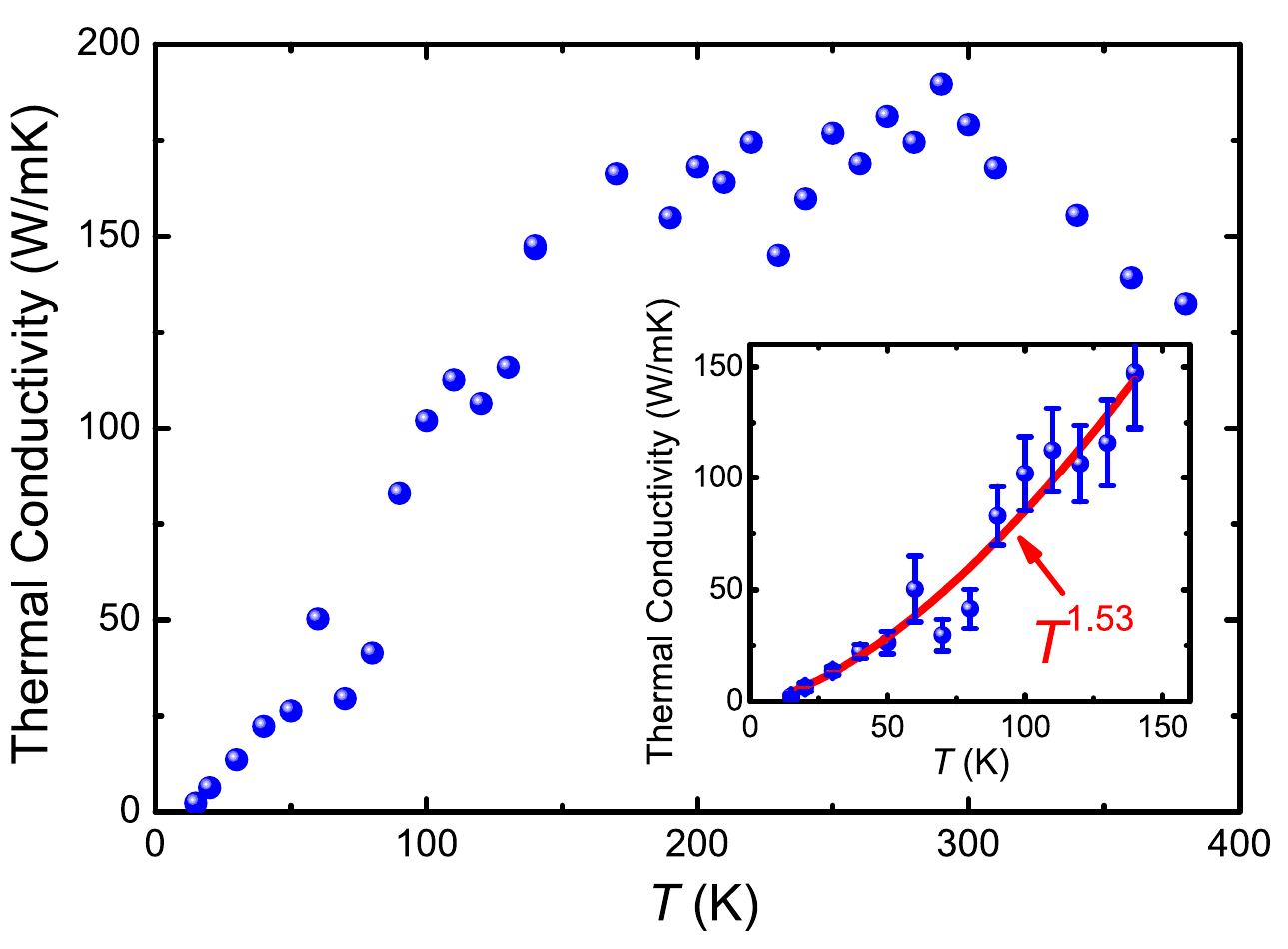}
\caption{Thermal conductivity $\kappa$ versus temperature. Inset:
thermal conductivity at low temperatures; the data can be fitted
by $\kappa$$=$b$T^{n}$, where b is the fitting parameter and $n$
is found to be 1.53($\pm$0.18).}
\end{figure}

We now turn to the experimental results and discuss first the
temperature dependent thermal conductivity $\kappa$($T$). Figure 2
shows the thermal conductivity $\kappa$ in the temperature range
from 15K to 380K of the suspended sample G1. $\kappa$ is extracted
from $\kappa$$=$$\sigma$L$/$A, where L is the sample length, A is
the cross section area of graphene and $\sigma$ is the measured
thermal conductance of graphene \cite{supplementary}. The observed
$\kappa$ increases by two orders of magnitude as the temperature
increases, reaches a maximum of 190 W/mK at $T\sim$280K($\pm$10K)
and decreases at higher temperature. The most important result is
the low temperature behavior of $\kappa$. The inset of Figure 2
shows that $\kappa$ can be fitted by $\kappa=$b$T^{n}$. Here, b is
the fitting parameter and $n$ is found to be 1.53($\pm$0.18).
Theoretical work has predicted that the thermal conductivity in
graphene yields a power law of $\sim$$T^n$, where $n$ varies from
1 for one-dimensional nanoribbons to 1.5 for two-dimensional
graphene sheets \cite{ballisticNL}. It is worth noting that this
$T^{1.5}$ power law at low temperatures was not observed in
supported graphene \cite{LSScience}. Our measurements of the
$T^{1.5}$ law directly reveals for the first time unambiguously
intrinsic phonon transport in graphene sheets.

We shall now interpret the physics behind the observed
$\kappa$($T$). The acoustic vibrations in 2D graphene lattice are
composed of two types of phonons: in-plane phonons (TA and LA
phonons) with linear dispersion, and out-of-plane phonons (ZA
phonons or flexural acoustic phonons) with quadratic dispersion.
It was proposed that only in-plane acoustic phonons carry heat in
graphene and the contribution from out-of-plane phonons can be
neglected \cite{BalandinTheory}. This arises from the fact that
the group velocity of ZA phonons is approaching zero for wave
vector, \textbf{q}$\rightarrow$0. However, Mingo $et$ $al$. have
argued that the ZA phonons carry most of the heat in SLG
\cite{ZAphonon}. This is due to the large density of modes for ZA
phonons resulting from their quadratic dispersion. At low
temperatures, the in-plane LA and TA modes would introduce a
$\sim$$T^{2}$ contribution to the thermal conductivity, while the
contribution from ZA mode scales as $\sim$$T^{1.5}$
\cite{ballisticMingo,ballisticNL}. The observed $\sim$$T^{1.5}$
behavior therefore proves that the thermal conductivity in
suspended SLG mainly arises from ZA phonons.

Interestingly, even before the first exfoliation of graphene,
Klemens \cite{Klemens} had predicted that low-frequency phonon
leakage into the substrate will reduce thermal conductivity by
20$\%$ to 50$\%$. Therefore, the measured $\kappa$(T) in this work
is much closer to the intrinsic phonon transport than that
measured in supported graphene. The observed $T^{1.5}$ behavior is
different from that of supported graphene and suggests that ZA
phonons carry most of the heat in suspended SLG. Theoretical
calculations also show that ZA phonons contribute more than 75$\%$
to $\kappa$ in suspended graphene below room tempereture
\cite{ZAphonon}, which is consistent with our results. The same
power law is also observed in samples with much inferior quality,
suggesting that this $T^{1.5}$ behavior is not strongly dependent
on the quality of the samples and may be intrinsic to phonon
transport in 2D systems (See Supplementary Information).

\begin{figure}[tbp]
\includegraphics[width=7.5cm]{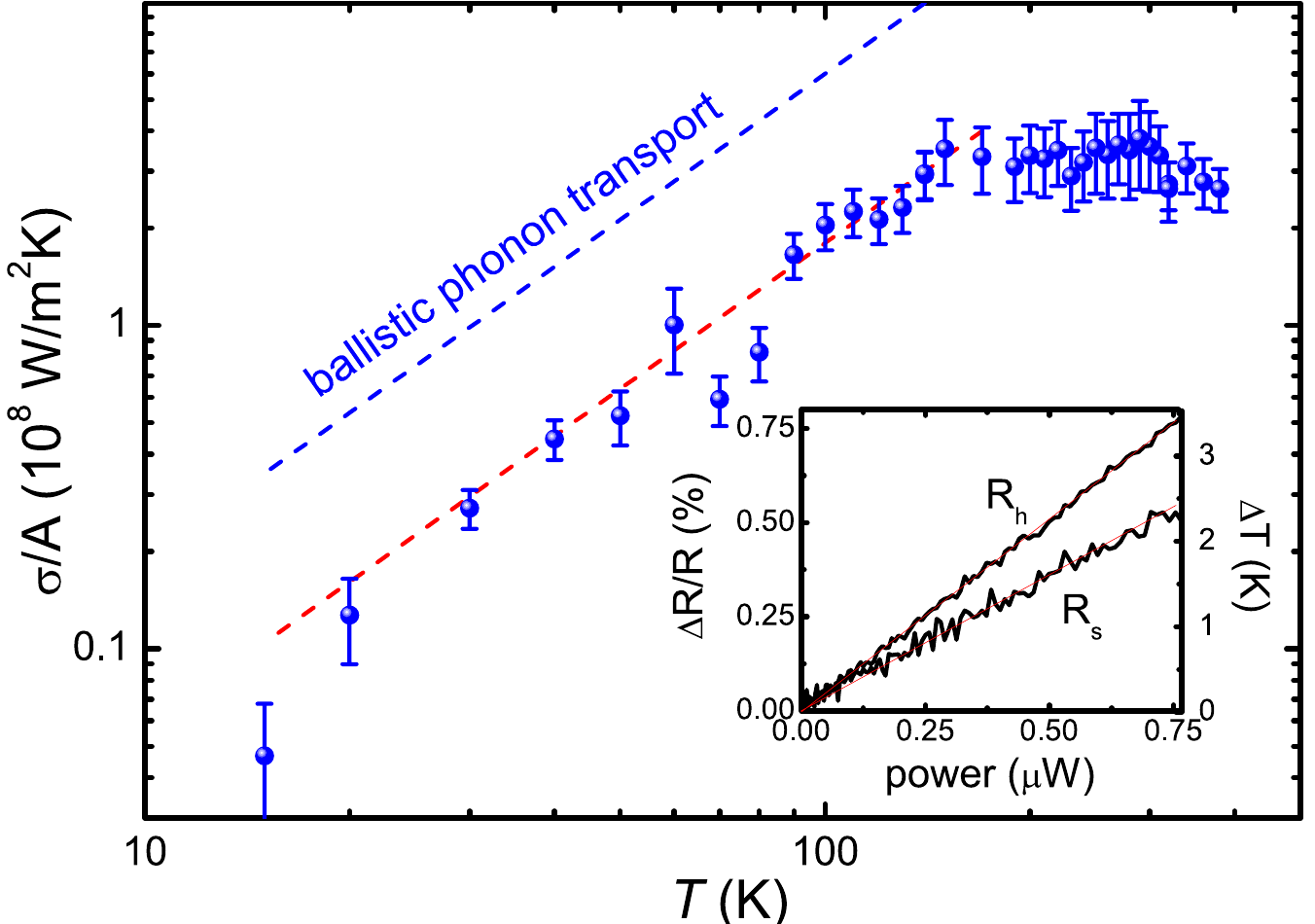}
\caption{Thermal conductance per unit cross section $\sigma$/A of
sample G1 as a function of temperature (solid circles). The data
can be fitted by 1.7$\times$10$^{5}$$T$$^{1.53}$ W/m$^{2}$K (red
dashed curve). The measured data is approaching the expected
ballistic limit in graphene sheets (blue dashed curve). Inset: the
temperature and resistance changes in the heater resistor R$_{h}$
and sensor resistor R$_{s}$ as a function of applied Joule heat,
the red lines are guides to the eye.}
\end{figure}

As the temperature increases further, the strong phonon-phonon
Umklapp scattering becomes more effective and dominates the
thermal conductivity due to thermally excited higher energy
phonons. It leads to a deviation from the $\sim$$T^{1.5}$ law and
a maximum in thermal conductivity around $\sim$280K ($\pm$10K),
followed by a decrease with a further increase of temperature.
This peak in thermal conductivity near room temperature is
consistent with experimental reports in supported graphene
\cite{LSScience,Lau} and carbon nanotubes \cite{Nanotube}. At low
temperatures where the phonon-phonon Umklapp scattering is
suppressed, phonon transport will approach the ballistic limit if
the graphene sheet is clean enough.

Next, we discuss the thermal conductance per unit cross section
$\sigma$/A as a function of temperature. At room temperature
$\sigma$/A is 3.8$\times$10$^{8}$ W/m$^{2}$K (Figure 3). This
value is comparable to the values observed in suspended exfoliated
SLG samples based on Raman measurements (3-5$\times$10$^{8}$
W/m$^{2}$K) \cite{balandin2,balandin3}. The high quality of CVD
graphene is further confirmed by the temperature dependence of
$\sigma$/A. It decreases by a factor of 100 to approximately
4.5$\times$10$^{6}$ W/m$^{2}$K at 15 K. Such a steep decrease has
not been observed previously, but more interestingly, the data at
low temperatures can be fitted with
1.7$\times$10$^{5}$$T$$^{1.53}$ W/m$^{2}$K. This is shown as the
dashed red curve in Figure 3 and has important implications for
the nature of phonon transport. Mingo and Brodio proposed that for
ballistic phonon transport, $\sigma$/A follows
6$\times$10$^{5}$$T$$^{1.5}$ W/m$^{2}$K \cite{ballisticMingo}, as
shown by the blue dashed curve in the same figure. The
experimentally observed value is close to 30$\%$ of the predicted
ballistic thermal conductance for graphene. From this we conclude
that in our sample with a channel length of 500nm phonon transport
approaches the ballistic limit. It is important to note that even
in multiwalled CNT, the experimentally observed $\sigma$/A reached
at most 40$\%$ of the ballistic limit in graphite
\cite{Nanotube,ballisticMingo}. Deviation from ballistic transport
in our experiment is most likely caused by scattering from PMMA
residues, CVD graphene specific organic contamination, ripples and
CVD graphene specific defects \cite{defectINCVD} such as wrinkles.
However, even in the absence of defects and contamination, two
additional source remain: a) scattering from naturally existing
1.1$\%$ $^{13}$C isotopic impurities \cite{isotopic} and b)
scattering from the boundaries between the suspended area and the
supported areas on top of the contacts.

\begin{figure}[tbp]
\includegraphics[width=7.5cm]{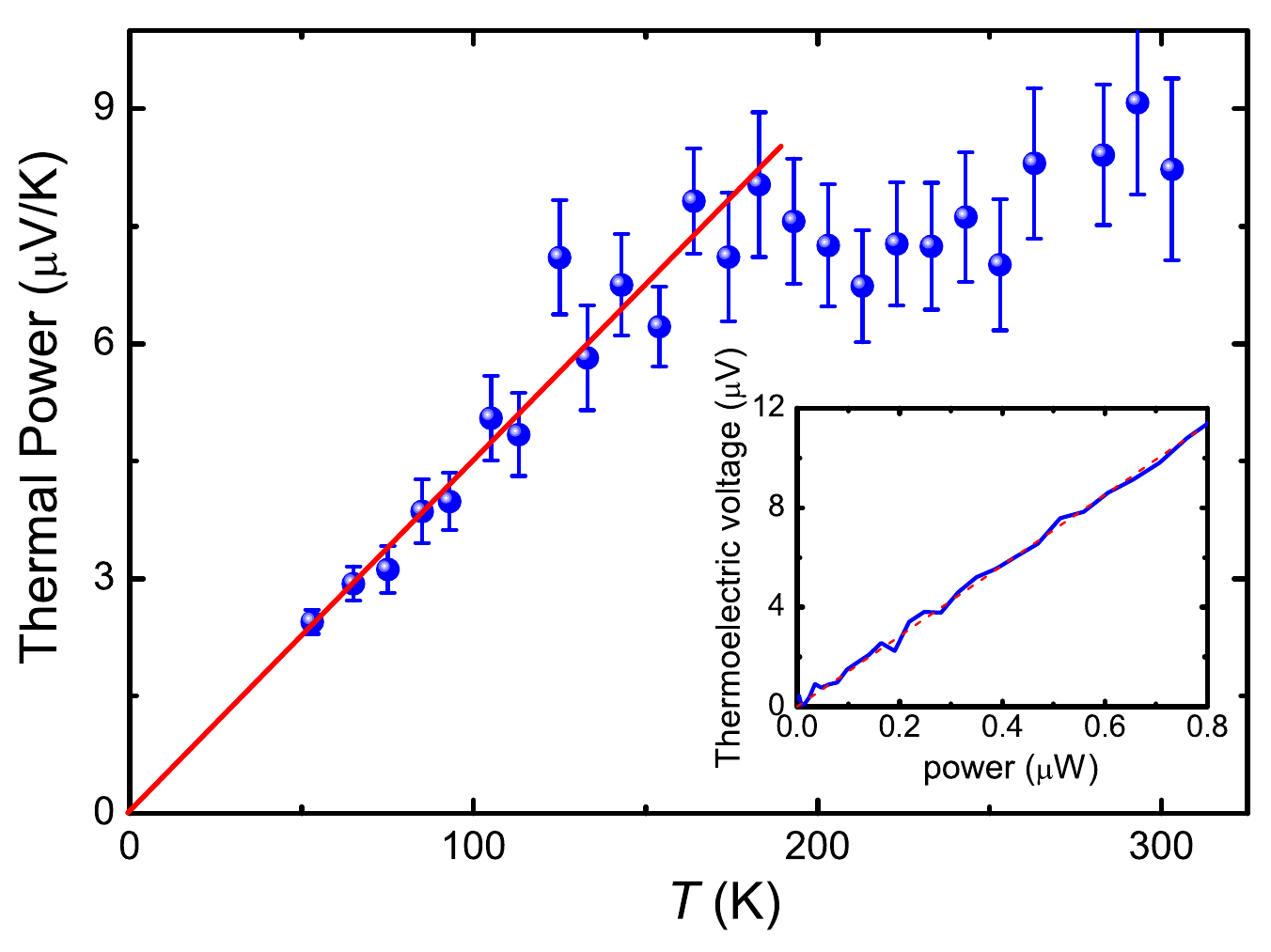}
\caption{The measured thermoelectric power (solid circles) and a
linear fit (red curve) for $T<$180K. Inset: the thermoelectric
voltage as a function of applied Joule heat, the red dashed line
is a guide to the eye.}
\end{figure}

Last but not least, we discuss the thermoelectric power (TEP) of
suspended graphene samples. Shown in the inset of Figure 4 is the
linear thermoelectric voltage as a function of Joule heat in
R$_{h}$ (blue curve). Since TEP and its temperature dependence
vary sharply across the Dirac point and we do not have a gate
control, a direct comparison with supported samples is difficult.
At room temperature the observed value is $+$9 $\mu$V/K,
suggesting a hole-like majority carrier. The linear fit (red curve
in Figure 4) to our data can be extrapolated to zero at $T$$=$0K.
Similar to TEP measurements in supported graphene
\cite{thermopower}, this confirms the absence of phonon drag in
suspended CVD SLG.

In conclusion we have measured phonon transport in suspended
graphene as a function of temperature from 380 K down to 15K. We
have taken advantage of the recent progress in Cu based CVD
graphene and observed a $T^{1.5}$ power in these samples. Our
results clearly show that the thermal conductance is strongly
dominated by ZA phonons. At low temperatures phonon transport is
approaching the ballistic limit, demonstrating the high quality of
CVD graphene samples also in the context of heat transport.

\begin{acknowledgments}
We thank Y. Zheng for helping with measurements and C.T. Toh for
helping with fabrication. This work is supported in part by
Singapore National Research Foundation (NRF-RF2008-07), NRF-CRP
grant (R-143-000-360-281), by a NUS grant (R-144-000-222-646), by
A*STAR grant (R-143-000-360-311), and by NUS NanoCore.
\end{acknowledgments}


\begin{thebibliography}{0}
\bibitem{lowD}A. Dhar, Adv. Phys. \textbf{57}, 457 (2008); L. Wang
and B. Li, Phys. World \textbf{21}, 27 (2008).
\bibitem{graphene}K.S. Novoselov $et$ $al$., Science \textbf{306}, 666
(2004); K.S. Novoselov $et$ $al$., Nature \textbf{438}, 197
(2005); Y.B. Zhang, Y.W. Tan, H.L. Stormer and P. Kim, Nature
\textbf{438}, 201 (2005).

\bibitem{neto}A.H. Castro Neto $et$ $al$., Rev. Mod. Phys. \textbf{81}, 109
(2009).
\bibitem{pop}E. Pop,
Nano Res. \textbf{3}, 147 (2010).
\bibitem{balandin2}A.A. Balandin $et$ $al$.,
Nano Lett. \textbf{8}, 902 (2008).


\bibitem{balandin1}S. Ghosh $et$ $al$., Appl. Phys. Lett. \textbf{92}, 151911 (2008).
\bibitem{balandin3} S. Ghosh $et$ $al$., Nat. Mater. \textbf{9}, 555 (2010).
\bibitem{diamond}L. Wei $et$ $al$., Phys. Rev. Lett. \textbf{70}, 3764 (1993).

\bibitem{Nanotube}P. Kim, L. Shi, A. Majumdar and P.L. McEuen, Phys. Rev. Lett. \textbf{87}, 215502 (2001).
\bibitem{CNT}E. Pop $et$ $al$., Nano Lett. \textbf{6}, 96 (2006).
\bibitem{Geim}C. Faugeras $et$ $al$., Acs Nano \textbf{4}, 1889 (2010).
\bibitem{Ruoff}W.W. Cai $et$ $al$., Nano Lett. \textbf{10}, 1645 (2010).
\bibitem{LSScience}J.H. Seol $et$ $al$., Science \textbf{328}, 213 (2010).

\bibitem{CVDruoff}X.S. Li $et$ $al$., Science \textbf{324},
1312 (2009).
\bibitem{CVDKorea}S. Bae $et$ $al$., Nat. Nanotechnol. \textbf{5},
574 (2010).


\bibitem{ballisticMingo}N. Mingo and D.A. Broido, Phys. Rev. Lett. \textbf{95}, 096105 (2005).
\bibitem{ballisticNL}E. Mu\~noz, J.X. Lu and B.I. Yakobson, Nano Lett. \textbf{10}, 1652 (2010).

\bibitem{Nanotube2}C.W. Chang, D. Okawa, A. Majumdar and A. Zettl, Science \textbf{314}, 1121 (2006).
\bibitem{Nanowire}R.K. Chen $et$ $al$., Phys. Rev. Lett. \textbf{101}, 105501
(2008); A.I. Hochbaum $et$ $al$., Nature  \textbf{451}, 163
(2008).
\bibitem{JHT}L. Shi $et$ $al$., J. Heat Transfer \textbf{125}, 881 (2003).
\bibitem{grainsize}The grain size of CVD SLGs is in the order of
20$\mu$m$\times$20$\mu$m, while the devices for the thermal
measurements have lateral dimensions of
$\approx$3$\mu$m$\times$1$\mu$m. Hence, the samples have, if at
all only one grain boundary at most.
\bibitem{supplementary}please refer to the supplementary informaion for details.
\bibitem{suspended}X. Du, I. Skachko, A. Barker and E.Y. Andrei, Nat. Nanotechnol. \textbf{3}, 491 (2008).
\bibitem{BalandinTheory}D.L. Nika, E.P. Pokatilov, A.S. Askerov and A.A. Balandin, Phys. Rev. B \textbf{79}, 155413
(2009).
\bibitem{ZAphonon}L. Lindsay, D.A. Broido and N. Mingo, Phys. Rev. B \textbf{82}, 115427 (2010).

\bibitem{Klemens}P.G. Klemens, Int. J. Thermophys., \textbf{22}, 265
(2001).
\bibitem{Lau}W.Y. Jang $et$ $al$., Nano Lett. \textbf{10}, 3909 (2010).

\bibitem{defectINCVD}O.V. Yazyev and S.G. Louie, Nat. Mater. \textbf{9}, 806
(2010).

\bibitem{isotopic}J.W. Jiang, J.H. Lan, J.S. Wang and B. Li, J. Appl. Phys. \textbf{107}, 054314
(2010).

\bibitem{thermopower}Y.M. Zuev, W. Chang and P. Kim, Phys. Rev. Lett. \textbf{102}, 096807
(2009); J.G. Checkelsky and N.P. Ong, Phys. Rev. B \textbf{80},
081413(R) (2009).





\end{thebibliography}
\end{document}